\documentclass[prc,onecolumn,english,showpacs]{revtex4}
\usepackage[T1]{fontenc}
\usepackage[latin1]{inputenc}
\usepackage{babel}
\usepackage{graphics}
\usepackage{comment}

\makeatletter

\providecommand{\LyX}{L\kern-.1667em\lower.25em\hbox{Y}\kern-.125emX\@}
\let\SF@@footnote\footnote
\def\footnote{\ifx\protect\@type
set@protect
    \expandafter\SF@@footnote
  \else
    \expandafter\SF@gobble@opt
  \fi
}
\expandafter\def\csname SF@gobble@opt \endcsname{\@ifnextchar[
  \SF@gobble@twobracket
  \@gobble
}
\edef\SF@gobble@opt{\noexpand\protect
  \expandafter\noexpand\csname SF@gobble@opt \endcsname}
\def\SF@gobble@twobracket[#1]#2{}
\def\nuc#1#2{\relax\ifmmode{}^{#1}{\protect\text{#2}}\else${}^{#1}$#2\fi}



\makeatother
\begin{document}

\title{Are spectroscopic factors from transfer reactions consistent with asymptotic normalisation
coefficients?}
\author{D.Y. Pang}
\affiliation{School of Physics and MOE Key Laboratory of Heavy Ion
Physics, Peking University, Beijing, China}
\affiliation{N.S.C.L. and Department of Physics and Astronomy,
Michigan State University, East Lansing, MI 48864, U.S.A.}
\author{F.M. Nunes}\email{nunes@nscl.msu.edu}
\affiliation{N.S.C.L. and Department of Physics and Astronomy,
Michigan State University, East Lansing, MI 48864, U.S.A.}
\author{A.M. Mukhamedzhanov}\email{akram@comp.tamu.edu}
\affiliation{Cyclotron Institute, Texas A\& M University,
College Station, TX 77843, U.S.A.}


\begin{abstract}
It is extremely important to devise a reliable method to extract spectroscopic factors
from transfer cross sections.
We analyse the standard DWBA procedure and combine it
with the asymptotic normalisation
coefficient, extracted from an independent data set.
We find that the single particle parameters
used in the past generate inconsistent asymptotic
normalization coefficients. In order to obtain a
consistent spectroscopic factor, non-standard parameters
for the single particle overlap functions can be used but,
as a consequence, often reduced spectroscopic strengths emerge.
Different choices of optical potentials and  higher order effects
in the reaction model are also studied. Our test cases consist
of:  $^{14}$C(d,p)$^{15}$C(g.s.) at $E_d^{lab}=14$ MeV,
$^{16}$O(d,p)$^{17}$O(g.s.) at $E_d^{lab}=15$ MeV and
$^{40}$Ca(d,p)$^{41}$Ca(g.s.) at $E_d^{lab}=11$ MeV.
We underline the importance of performing experiments specifically
designed to extract ANCs for these systems.
\end{abstract}
\maketitle

\section{Introduction}


The shell model formalism first introduced Spectroscopic Factors (SF)
to describe the shell occupancy \cite{brown}.
In particular, $S^{f,i}_{lj}$, the single particle SF, is
defined as the norm of the overlap function of a nucleus (A+1)
in a particular state $i$ with a nucleus A in a state $f$,
where the valence nucleon
is in an orbital with orbital and total angular momentum $(l,j)$.
These SFs have been extensively compared
with those extracted from  reactions.
At present ab-initio calculations are improving the
accuracy of the calculated SFs (e.g. \cite{gfmc,gfmc-eep}).
One would eventually like to have a very accurate probe that could test
the predictions of these models and could disentangle the
relevant elements of the NN force that are still missing,
especially when moving toward the driplines.


From the experimentalist point of view, a spectroscopic factor is
a ratio of measured to predicted differential cross sections.
Phenomenological spectroscopic factors are extensively
used in a variety of topics, from nuclear reactions to astrophysics or
applied physics,  yet the procedure for
their extraction from the data has remained essentially the same for
decades. Most of the work in the sixties and seventies used direct
transfer reactions, such as
$(d,p),\,(d,t),\,({}^{3}{\rm He},d),\,({}^{3}{\rm He},\alpha)$,
as the central tool \cite{goncharov,austern70,jpg-rev}.
However transfer analyses have a reputation of large uncertainties.


Other methods to extract spectroscopic information
include the (e,e'p) reactions or the nuclear knockout
in inverse kinematics.
For stable nuclei, there have been many electron knockout experiments.
They have provided SFs with rather small error bars \cite{kramer}.
The analysis of these measurements along
a wide mass range,  for  single particle states with expected
SFs close to unity, show an overall reduction of the SF ($\approx 0.6$).
The source of this reduction is not yet well understood \cite{barbieri}
although one expects NN short or long range correlations
which are not included in the present
day non-ab-initio shell model to contribute.

Nuclear Knockout in inverse kinematics using radioactive beams
is a new probe introduced at the NSCL \cite{knock}.
A systematic study of spectroscopy on a variety of nuclei, ranging from
the stability valley to the proton dripline, have been performed
using this technique. It is found that the measured spectroscopic factors
suffer from a reduction relative to the Shell Model predictions
and the reduction factor changes with binding energy \cite{gade}. Again, it is not
clear where this quenching \cite{brown} comes from. Of course, as (e,e'p),
this technique is only suitable for studying the single particle structure
of the ground states of nuclei.

The (e,e'p) cross section is sensitive to the structure all
the way to the inside of the nucleus, whereas the transfer is typically peripheral
and surface peaked. The knockout reactions with radioactive
beams have been  performed in a kinematical regime where the eikonal approximation
can be used, to simplify the reaction theory and
to reduce the reaction model uncertainties \cite{jpg-rev}. 
Nuclear knockout results \cite{brown} are in agreement with the
(e,e'p) SFs for the tested stable closed shell nuclei \cite{gregers}.
Transfer and (e,e'p) have also been compared \cite{kramer} for $^{48}$Ca.
Earlier transfer studies provided
a spectroscopic factor for the last proton close to unity,
but a reanalysis by \cite{kramer} including
finite-range and non-local effects in the transfer reaction model,
show that the transfer result can be brought down to 0.6,
the (e,e'p) value.

The spectroscopic factor is the norm of the overlap function,
which peaks well inside the nuclear radius $R_N$.
As the (e,e'p) reaction can probe the nuclear interior, it is suitable
for  extracting SFs as long as the reaction mechanism is well understood.
However there are some problems with the high momentum
transfer component, associated with probing the inside of the nucleus,
because then the Born approximation is not valid \cite{born}.
For exotic nuclei near or on the driplines,
transfer reactions are a unique tool and, hence, can have a large
impact in the programs of the new generation rare isotope laboratories.


The standard framework for analysing transfer data
with the intent of extracting SFs is the distorted-wave Born approximation
(DWBA). Overall, it has been very successful in describing angular distributions
at forward angles and less so for the larger angles where higher order become
more important. The SF is the normalization needed for the calculated DWBA
differential cross section to match the experimental one at forward angles
(e.g. \cite{schiffer,dwba,iliadis04}).
The uncertainty of the extracted SF resulting from the normalization of the DWBA cross section
is assumed to be  $\sim 30$\%, even if the statistical errors are low.
The reasons for this inaccuracy are typically attributed to ambiguities in the
optical potentials, the inadequacy of the DWBA reaction theory, or
the dependence on the single-particle potential parameters.
Recently, systematic studies on $^{12}$C(d,p)$^{13}$C have shown that
it is possible to bring SFs into conformity  using a global optical
potential prescription \cite{liu,tsang}, whereas arbitrary choices of the
optical potential will hold disparate results.
As to the reaction mechanism, there are many studies on the validity of DWBA
(e.g. \cite{delaunay}) and typically the reaction mechanism needs to be checked
case by case. In addition, Hartree-Fock densities have been suggested as a mean
to constrain the single particle parameters \cite{jenny}. Therein, the single particle
radii for the Ca isotopes were adjusted to reproduce the known rms matter radii,
and zero range DWBA calculations were performed to extract the SFs.

In \cite{us}, a combined method of extracting SFs from transfer reactions was introduced.
This method can also be applied to breakup and (e,e'p) reactions.
The combined method, which is based on the introduction of the
asymptotic normalization coefficients (ANC) into the transfer analysis, allows one to significantly
reduce the uncertainty in the choice of the bound state potential parameters and
to test the DWBA or other underlying reaction theory. In the combined method
the ANC should be determined from an independent measurement of a
peripheral reaction while the SF is determined from transfer reactions
which are sensitive to the nuclear interior. In \cite{us} we emphasize that fixing the ANC
is absolutely necessary, since even when the beam energy is well above the Coulomb
barrier, most of the reaction happens in the asymptotic region.
It has been found \cite{us} that, in $(d,p)$ reactions, the standard single particle parameters for the radius 
$r_0=1.2$ fm and diffuseness $a=0.65$, which typically provide unit SF for closed shell
nuclei, do not reproduce the value of the ANC extracted from
the lower energy measurement. In this work we expand on the ideas of the
combined method, and  explore
other uncertainties (such as optical potentials and higher order effects)
to attempt a unification of the SF and the ANC, searching for a reaction
description which is practical and gives reliable spectroscopic information.


In section II we present a short description of the
theoretical framework. In Section III we detail the results for our three test cases:
IIIA for DWBA results using a global deuteron optical potential,
IIIB for DWBA results where the deuteron potential is obtain from a direct
fit to the elastic, IIIC for results including deuteron breakup within the adiabatic model,
IIID for checks of other higher order effects and  also in 
IIIE a further analysis of peripherality.
Finally,  a discussion of the results and conclusions
are drawn in section IV.

\section{Reaction formalism and spectroscopy}

The central element of the analysis of the transfer reaction $A(d,p)B$ \cite{us} is the overlap
function  $I^{B}_{An}({\rm {\bf r}})$ between
bound-states of nuclei $B=A+n$ and $A$ which
depends on  $\,{\bf {\rm \bf r}}$, the radius-vector connecting the
center of mass of $A$ with $n$.
The square norm of the overlap function gives a model-independent
definition of the SF. It is important to keep in mind that most of
the contribution to the SF comes from the interior.

The  radial overlap function (for $B= A +n$) behaves as a spherical Bessel
function for large distances:
\begin{equation}
I^{B}_{An(lj)}(r) \stackrel{r > R_{N}}{\approx}
C_{lj}\,i\,\kappa\,h_{l}(i\,\kappa\,r).
\label{ovasymptotic}
\end{equation}
Here, $\kappa= \sqrt{2\,\mu_{An}\,\varepsilon_{An}}$, where
$\varepsilon_{An}$
is the binding energy for $B \to A + n$, and $\,\mu_{An}$
is the reduced mass of $A$ and $n$.
$R_{N}$ represents a radius beyond which the nuclear potential is negligible and
$C_{lj}$ is referred to as the asymptotic normalization coefficient (ANC).

The standard practice is to take the radial dependence
of the overlap function from the single particle orbital.
This radial wavefunction is usually generated by a Woods Saxon potential
with a given geometry. The depths are adjusted such that the single particle orbital
has the correct separation energy and quantum numbers ($n,l,j$).
Generally, the single particle orbital has the same asymptotic behaviour as
the many-body overlap function:
\begin{equation}
\varphi_{An(n_{r}lj)}(r)\stackrel{r > R_{N}}{\approx}
b_{n_{r}lj}\,i\,\kappa\,h_{l}(i\,\kappa\,r),
\label{spasymptotic}
\end{equation}
where $b_{n_{r}lj}$ is the single-particle ANC (SPANC).
From the relations in Eqs. (\ref{ovasymptotic}) and (\ref{spasymptotic}),
and under the assumption that the many-body overlap is indeed
proportional to the single particle function all the way down to $r=0$,
the ANC and the SF are related by  $C^2_{lj}= S_{n_{r}lj}\,b^2_{n_{r}lj}$.
So far, it is hard to check the exactness of this proportionality
given the accuracy of the asymptotics of the wavefunction
obtained from ab-initio calculations.

For  $A(d,p)B$, the one-step post-form finite range DWBA amplitude is given by:
\begin{equation}
M= <\psi_{f}^{(-)}I^{B}_{An}|\Delta V|\varphi_{pn}\,\psi_{i}^{(+)}>,
\label{dwba1}
\end{equation}
where $\Delta V= V_{pn} + V_{pA} - U_{pB}$ is the transition
operator,
$V_{ij}$ is the interaction potential between $i$ and $j$, and
$U_{pB}$ is the optical potential in the final-state.
$\psi^{(+)}_{i}$ and $\psi^{(-)}_{f}$ are the distorted waves in
the initial and final states and $\varphi_{pn}$ is the deuteron bound-state
wave function.

In \cite{us},  the reaction amplitude is split
into interior  and  exterior parts.
The normalisation of the former is determined by the SF while the ANC governs the normalisation
of the latter. We introduce the ratio $R^{th}(b_{n_rlj})=\frac{\sigma^{th}(\theta_{peak})}{b_{n_rlj}^2}$
to compare with the experimental counterpart
${\cal R}^{exp}=\frac{\sigma^{exp}(\theta_{peak})}{(C^{exp}_{lj})^2}$.
Introducing the information about the ANC fixes the exterior contribution.
If the reaction is completely peripheral, its cross section will scale directly
as $C^2_{lj}$, but will hold no information on the SF.
If there is an interior contribution, the theoretical cross section has a non-trivial
dependence on the SPANC $b_{n_rlj}$ which can be constrained by $R^{th}(b_{n_rlj})={\cal R}^{exp}$.
Hence, peripheral reactions
are ideal to extract ANCs but useless for extracting SFs.
To determine SFs one should explore non-peripheral reactions.

The combined method presented in \cite{us}  tries to isolate  the ambiguity
coming from the single particle parameters but the resulting SFs
are often smaller than those produced by shell model.
As in this method, the interior part plays an important role to determine the SF,
the results will be more sensitive to the optical potentials and coupling effects.
With global optical potentials, the single particle parameters obtained from
$R^{th}(b_{n_rlj})={\cal R}^{exp}$  are often far from the conventional values,
and the corresponding angular distributions provide a worse description of
the data, when compared to the standard procedure. The main questions we want
to address here is whether DWBA (or higher order reaction theory) allows us to extract
correct SFs when fixing the peripheral part of the transfer amplitude through
the experimentally determined ANCs.

\section{Results}

The most important consequence of the work in \cite{us} is the fact that
within the standard DWBA approach, the extracted spectroscopic factors are inconsistent
with the ANCs obtained through independent measurements. 
This will be illustrated through the examples in this section.

By {\it standard DWBA} we mean the framework in which the 1-step transfer matrix
element is evaluated with incoming and outgoing distorted waves calculated
by fitting the deuteron and proton elastic scattering with local optical potentials.
The transfer operator contains finite range effects as well as the full complex remnant term.

\begin{figure}[t!]
\resizebox*{0.45\textwidth}{!}{\includegraphics{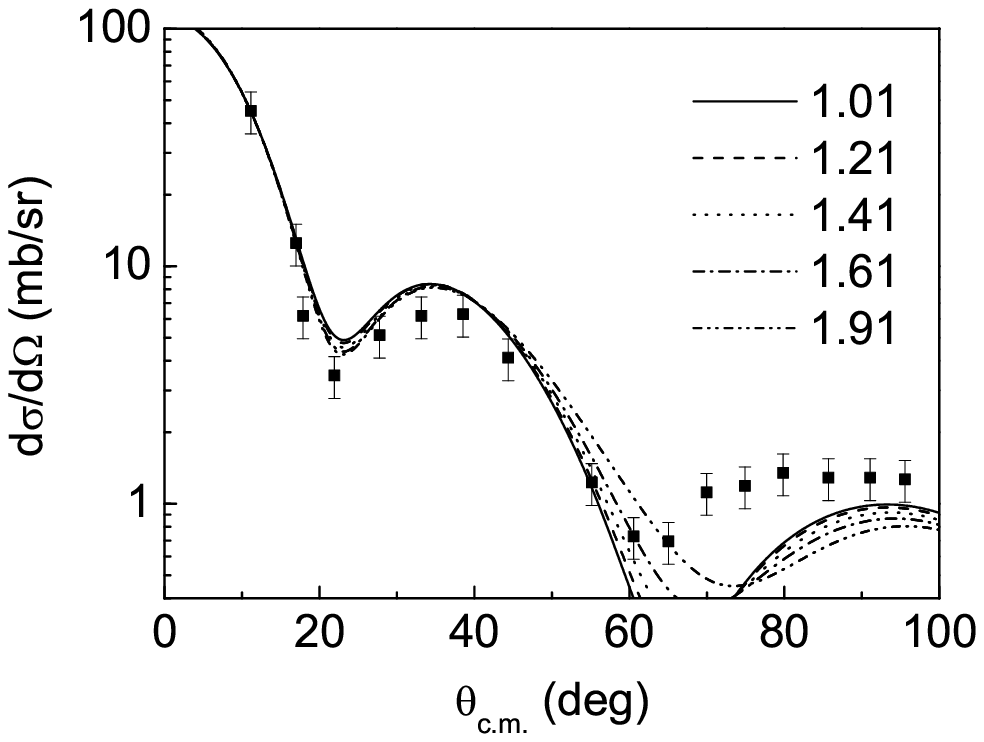}}
\vspace{-0.2cm}
\caption{\label{c15global}
$\nuc{14}{C}(d,p)\nuc{15}{C}$ at $E_d=14$ MeV with  a global deuteron potential (DWBAg).
Data from \cite{goss}.}
\resizebox*{0.45\textwidth}{!}{\includegraphics{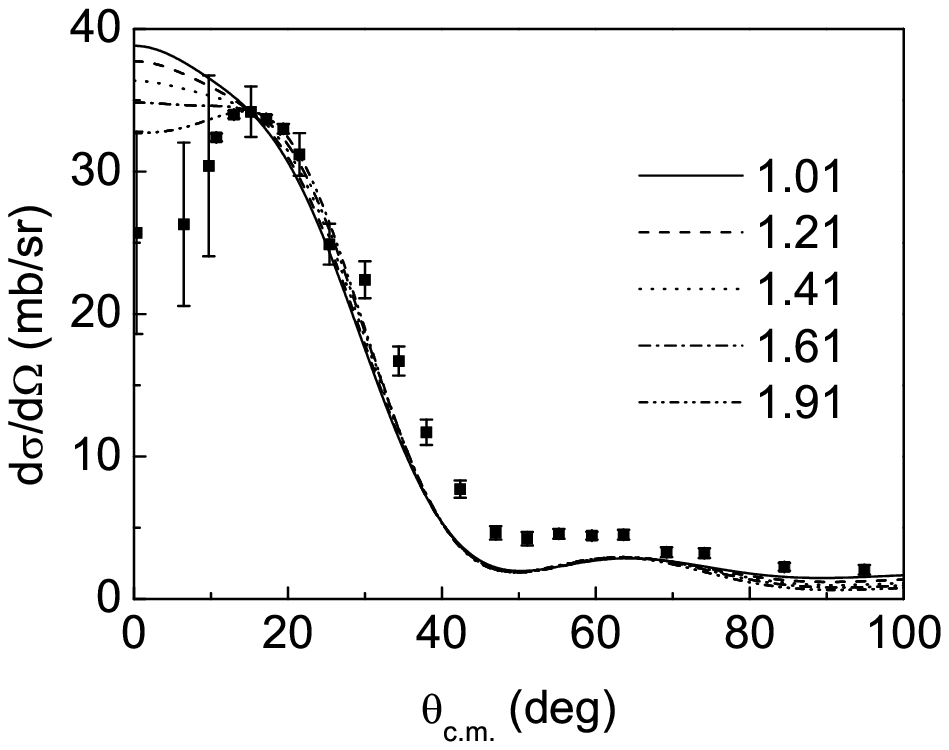}}
\vspace{-0.2cm}
\caption{\label{o17global}
$\nuc{16}{O}(d,p)\nuc{17}{O}$ at $E_d=15$ MeV with  a global deuteron potential (DWBAg).
Data from \cite{eldon}.}
\resizebox*{0.45\textwidth}{!}{\includegraphics{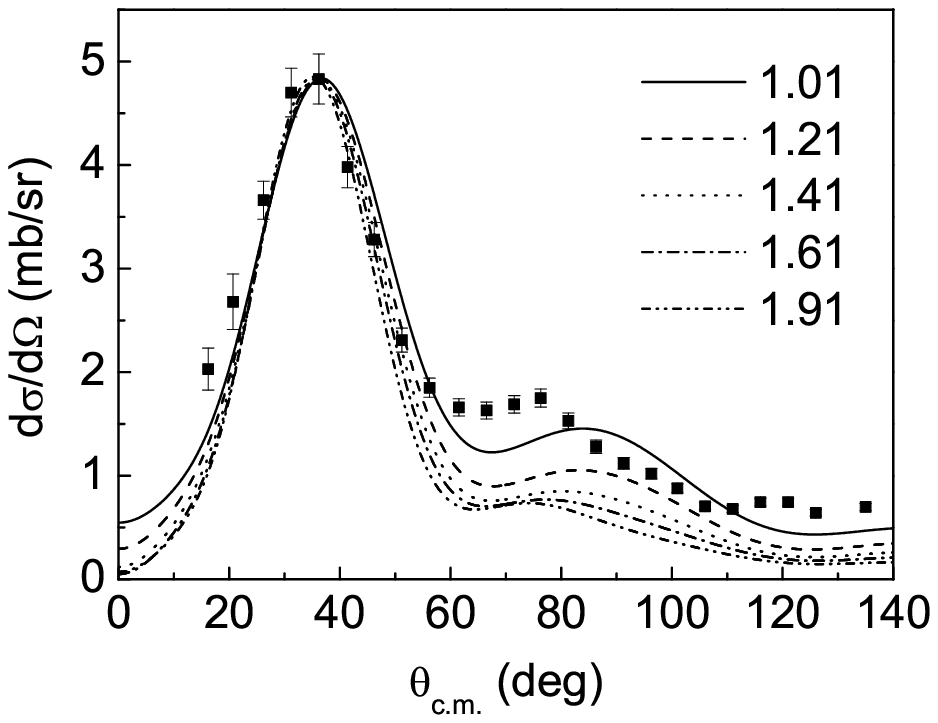}}
\vspace{-0.2cm}
\caption{\label{ca41global}
$\nuc{40}{Ca}(d,p)\nuc{41}{Ca}$ at $E_d=11$ MeV with  a global deuteron potential (DWBAg).
Data from \cite{kocher}.}
\end{figure}

ANCs can be extracted from sub-Coulomb reactions, which are
Coulomb dominated and contain virtually no contribution from the nuclear interior.
For the $^{17}$O(g.s.) case, we have two sets of heavy ion collision data  that provide
$C^2_{d5/2}=0.67 \pm 0.05$ fm$^{-1}$ \cite{o17anc1} and
$C^2_{d5/2}=0.69 \pm 0.03$ fm$^{-1}$ \cite{o17anc2} respectively.
For the other two cases, no heavy ion transfer data exist, therefore alternative
data was used.
For  $^{15}$C(g.s.) knockout data from \cite{sauvan} and \cite{maddalena}
was jointly used in \cite{c15anc} to extract an ANC of
$C^2_{s1/2}=1.48 \pm 0.18$ fm$^{-1}$. Note that other values can be obtained
through different types of reactions such as capture \cite{timofeyuk}. 
We want to be consistent with the ANC extractions in all three test cases, thus we use the 
ANC obtained through nuclear processes.
For $^{41}$Ca(g.s.), (d,p) data at sub-Coulomb
energies was used \cite{ca41anc} from which we obtain $C^2_{f7/2}=8.36 \pm 0.42$ fm$^{-1}$.

The transfer calculations are performed for $^{14}$C(d,p)$^{15}$C(g.s.) at $E_d^{lab}=14$ MeV
\cite{goss},
$^{16}$O(d,p)$^{17}$O(g.s.) at $E_d^{lab}=15$ MeV \cite{eldon}, and
$^{40}$Ca(d,p)$^{41}$Ca(g.s.) at $E_d^{lab}=11$ MeV \cite{kocher}. These reactions were
chosen because both elastic and transfer data exist at the same energy. All data can be found
on the database website at Michigan State University \cite{database}. Calculations
were performed with the code {\sc fresco} \cite{fresco}.

\subsection{DWBA with global deuteron potential}
\label{sec-dwbag}

We perform finite-range post-form DWBA calculations, 
including the full complex transition operator $\Delta V$.
The Reid-Soft-Core  interaction \cite{rsc} is used for the deuteron ground state wavefunction,
and includes both S- and D-waves. We use the central part of the  Reid-Soft-Core  interaction
for $V_{np}$ in the transfer operator. 
The diffuseness of the single particle orbital of the final state is kept fixed (a=0.65 fm),
but the radius is varied to generate a range of SPANCs. 
Throughout this work we will always use the CH89 global parameterization \cite{ch89} 
for the outgoing distorted waves. We have checked that by using a different proton global potential, 
spectroscopic factors change less than 10 \%. We will look into several approaches to determine 
the initial wavefunction. In this subsection we take the  Perey \& Perey deuteron potential (PP) 
\cite{perey}.

The resulting angular distributions
are displayed in Figs. \ref{c15global}, \ref{o17global} and \ref{ca41global},
for a subset of $r_0$, and compared to data. We will refer to these calculation as
DWBAg. Even though it provides a fair description of the first peak of the distribution,
it  is inadequate for the large angles, where higher order effects become important.
This is a well known characteristic of DWBA.

\begin{table}
 \caption{SFs and ANCs obtained from DWBA analyses: perey and perey potential for the deuteron.}
\begin{center}
\begin{tabular}{|c|c|c|c|c|c|c|c|c|c|}
 \hline
 &\multicolumn{3}{c|}{\nuc{14}{C} (14 MeV)}
 &\multicolumn{3}{c|}{\nuc{16}{O} (15 MeV)}
 & \multicolumn{3}{c|}{\nuc{40}{Ca} (11 MeV)} \\\hline
 $r_0$ (fm) & $b$ (fm$^{-1}$) &  SF  & $C^2$  (fm$^{-1}$)& $b$  (fm$^{-1}$)  &  SF  & $C^2$  (fm$^{-1}$)&
 $b$  (fm$^{-1}$) &  SF  & $C^2$  (fm$^{-1}$)\\ \hline
 1.01 & 1.342 &   1.40 & 2.516  &  0.675 &  1.54 & 0.700 & 1.322 &  1.66     & 2.900 \\
 1.11 & 1.377 &   1.33 & 2.527  &  0.753 &  1.28 & 0.723 & 1.664 &  1.18     & 3.273 \\
 1.21 & 1.415 &   1.27 & 2.541  &  0.841 &  1.05 & 0.745 & 2.091 & 0.834     & 3.647 \\
 1.31 & 1.454 &   1.21 & 2.554  &  0.940 & 0.869 & 0.767 & 2.623 & 0.583     & 4.018 \\
 1.41 & 1.496 &   1.15 & 2.572  &  1.050 & 0.716 & 0.789 & 3.283 & 0.407     & 4.390 \\
 1.51 & 1.540 &   1.09 & 2.592  &  1.173 & 0.589 & 0.811 & 4.099 & 0.284     & 4.775 \\
 1.61 & 1.586 &   1.04 & 2.621  &  1.310 & 0.486 & 0.833 & 5.107 & 0.199     & 5.198 \\
 1.71 & 1.635 &  0.995 & 2.659  &  1.462 & 0.401 & 0.857 & 6.347 & 0.141     & 5.682 \\
 1.81 & 1.686 &  0.953 & 2.710  &  1.631 & 0.332 & 0.883 & 7.872 & 0.101     & 6.260 \\
 1.91 & 1.740 &  0.918 & 2.777  &  1.818 & 0.275 & 0.911 & 9.744 & 0.073     & 6.968 \\ \hline
 & \multicolumn{2}{c|}{$C^2_{exp}$} & $1.48\pm0.18$ fm$^{-1}$ & \multicolumn{2}{c|}{$C^2_{exp}$}
 & $0.67\pm0.05$ fm$^{-1}$ & \multicolumn{2}{c|}{$C^2_{exp}$}  & $8.36\pm0.42$ fm$^{-1}$ \\
 \hline
 \end{tabular}
 \label{tab-global}
 \end{center}
 \end{table}

For each single particle radius $r_0$ of $B=A+n$, the normalization of the first transfer peak is
used to determine the spectroscopic factor, and the ANCs are obtained directly
from  $C^2_{lj}= S_{n_{r}lj}\,b^2_{n_{r}lj}$.
Results are presented
in Table \ref{tab-global}. If one takes the standard radius of $1.21$ fm, one can see
that the extracted SFs are all close to unity, but none of the ANCs are consistent with the
values extracted from experiment, with
the most serious mismatch for $^{41}$Ca, nearly a factor of 2.
Knockout measurements suggest that in $^{17}$O and $^{41}$Ca there is
a reduction of the SF to $\approx 0.6$. One can see from Table  \ref{tab-global},
that this SF is reproduced at $r_0 \approx 1.4$ fm for $^{17}$O and  $r_0 \approx 1.3$ fm for $^{41}$Ca,
but the SF/ANC inconsistency is not resolved.

Finite range effects are known to be important in (d,p) reactions, but remnant contributions
are often assumed to be small. We have repeated DWBAg calculations excluding the remnant
term and find that indeed, remnant contributions are insignificant for the reactions
of $^{14}$C and $^{16}$O, but can contribute  up to $30$\% for the $^{40}$Ca(d,p) reaction.

\subsection{DWBA with fitted deuteron potential}
\label{sec-dwbaf}

\begin{table}
\caption{Deuteron optical potential parameters resulting from the fit to elastic data (fit 1).}
\begin{center}
\begin{tabular}{|c|c|c|c|c|c|c|c|c|c|c|}
\hline
$A$& $E_{beam}$ (MeV) & $V_{R} $(MeV)  &  $R_{R}$ (fm)  & $a_{R}$ (fm) & $W_{d}$(MeV)
& $R_I$ (fm)  &  $a_I$ (fm)  & $V_{so}$(MeV) & $R_{so}$ (fm) & $a_{so}$ (fm)\\\hline
14 & 14 & 92.880 & 1.1486 & 0.596 & 3.848 & 1.085 & 1.347 & 9.164  & 0.883 &  0.244 \\
16 & 15 & 115.911 & 1.017 & 0.846 & 11.257 & 1.073 & 0.584 & 11.600  & 0.578 &  0.343 \\
40 & 11 & 115.177 & 1.040 & 0.712 &  5.287 & 1.375 & 0.856 &  4.486  & 0.382 &  0.266 \\\hline
\end{tabular}
\label{tab-optpot}
\end{center}
\end{table}

\begin{figure}[ht!]
\resizebox*{0.45\textwidth}{!}{\includegraphics{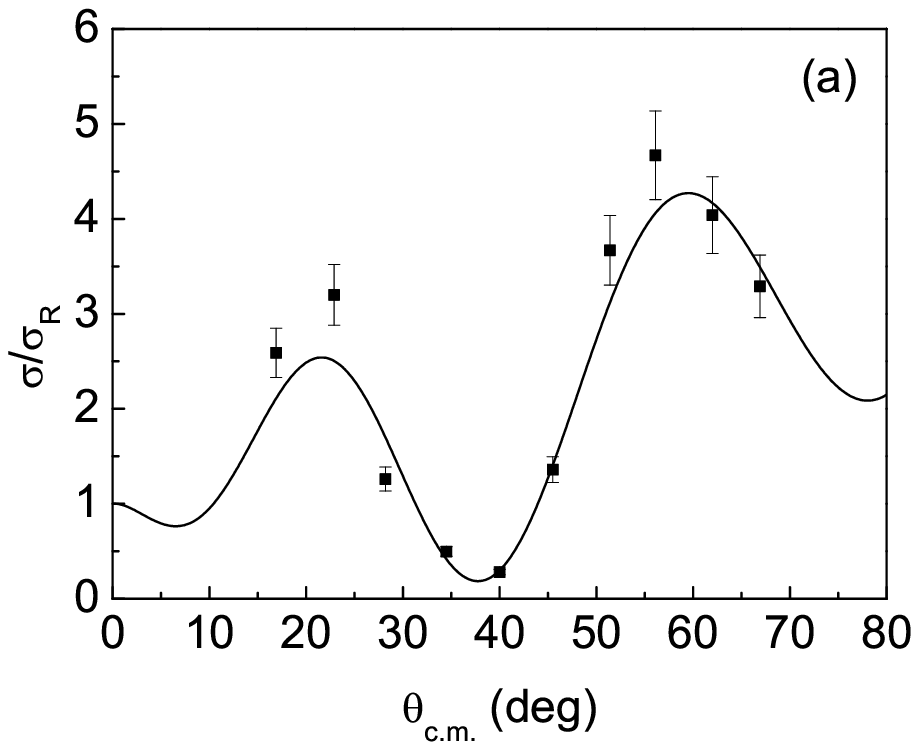}}
\resizebox*{0.45\textwidth}{!}{\includegraphics{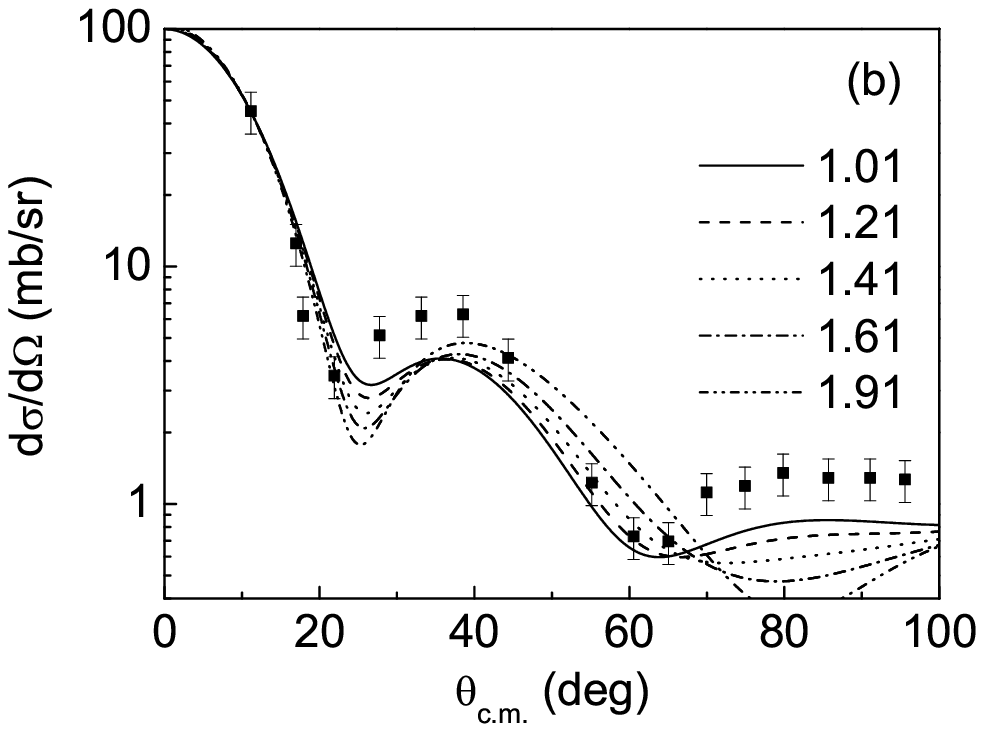}}
\vspace{-0.2cm}
\caption{\label{c15fitted}
a) Elastic scattering fit $\nuc{14}{C}+d$; b) $\nuc{14}{C}(d,p)\nuc{15}{C}$ at $E_d=14$ MeV with a fitted deuteron
potentials (DWBAf).
Both elastic and transfer data from \cite{goss}.}
\resizebox*{0.45\textwidth}{!}{\includegraphics{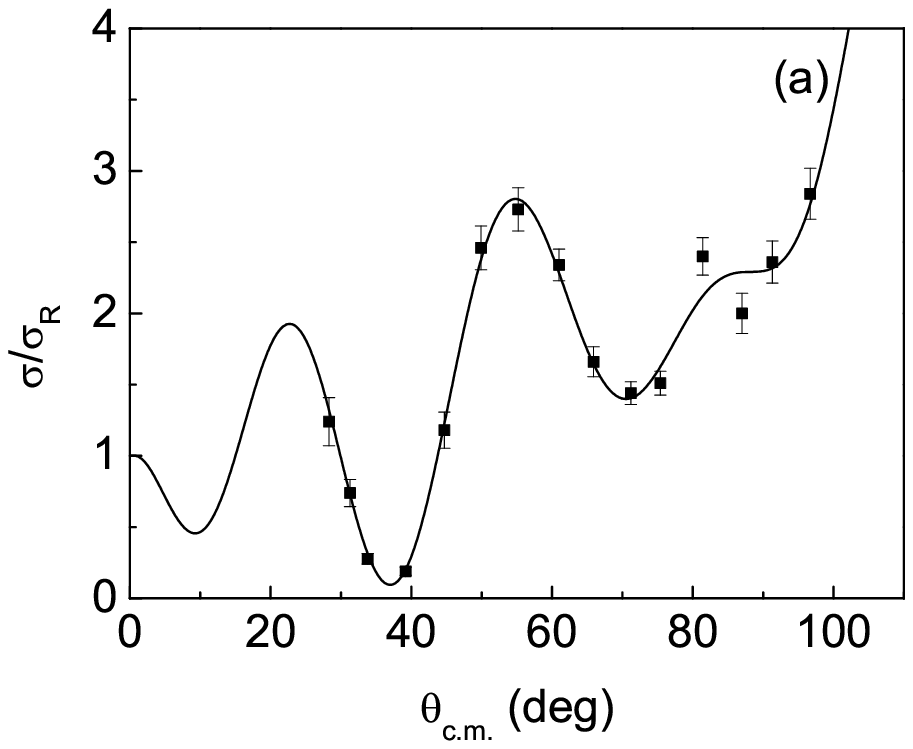}}
\resizebox*{0.45\textwidth}{!}{\includegraphics{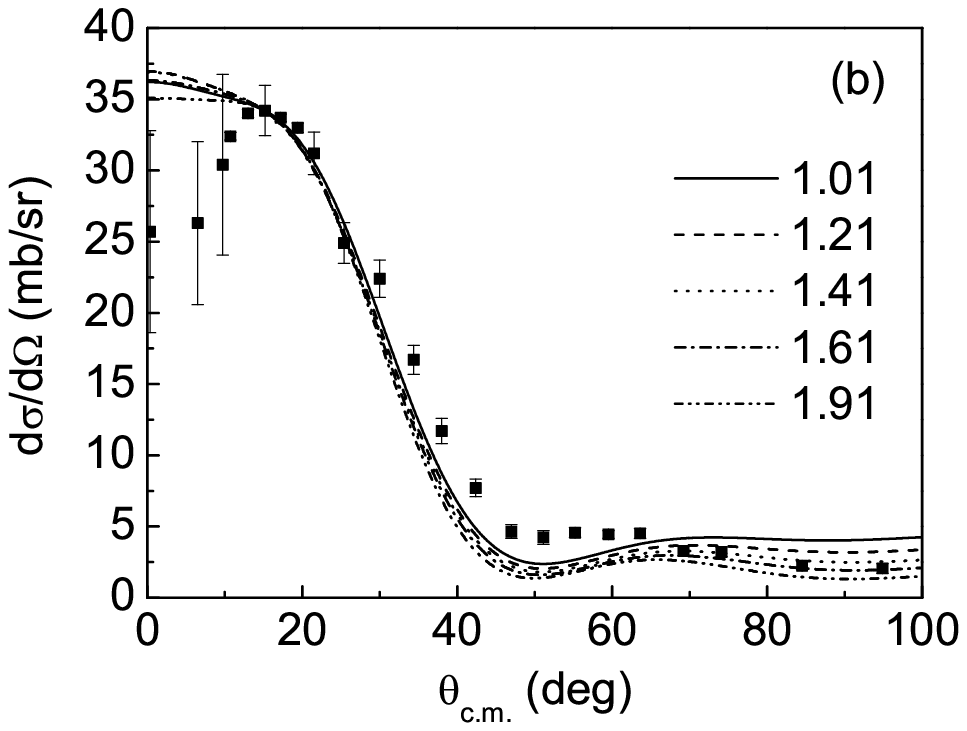}}
\vspace{-0.2cm}
\caption{\label{o17fitted}
a) Elastic scattering fit $\nuc{14}{C}+d$; b) $\nuc{16}{O}(d,p)\nuc{17}{O}$ at $E_d=15$ MeV with a fitted deuteron potentials
(DWBAf).
Elastic data from \cite{o16el} and transfer data from \cite{eldon}.}
\resizebox*{0.45\textwidth}{!}{\includegraphics{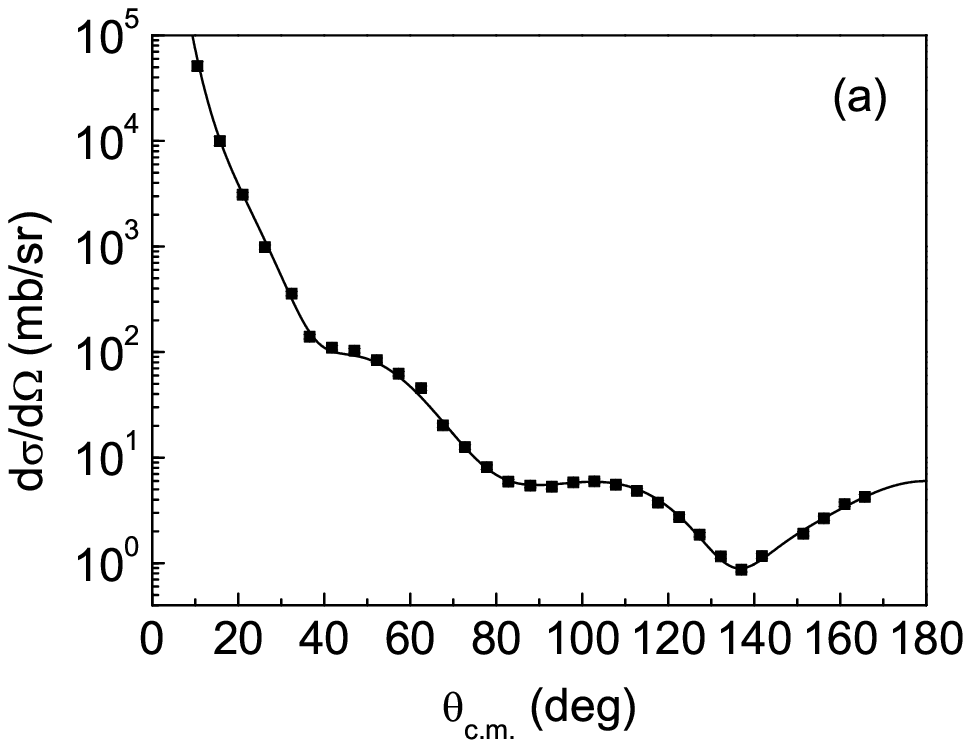}}
\resizebox*{0.45\textwidth}{!}{\includegraphics{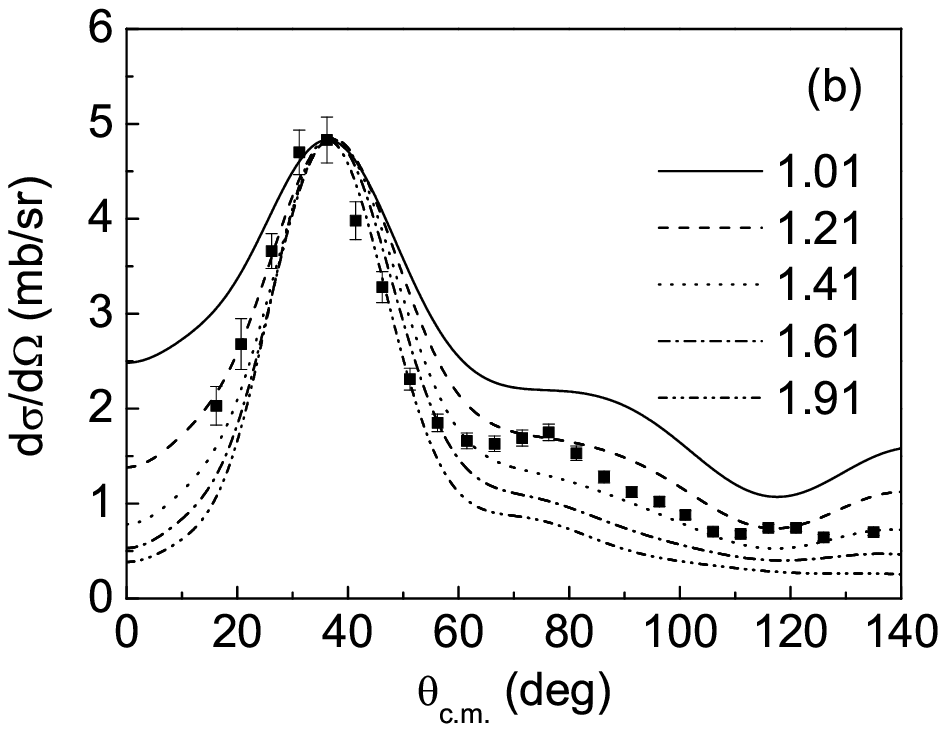}}
\vspace{-0.2cm}
\caption{\label{ca41fitted}
a) Elastic scattering fit $\nuc{14}{C}+d$; b) $\nuc{40}{Ca}(d,p)\nuc{41}{Ca}$ at $E_d=11$ MeV with a fitted deuteron potentials
(DWBAf).
Elastic data from \cite{ca40el} and transfer data from \cite{kocher}.}
\end{figure}

The elastic channel data is also available for the three reactions under scrutiny,
$^{14}$C(d,p)$^{15}$C(g.s.) at $E_d^{lab}=14$ MeV,
$^{16}$O(d,p)$^{17}$O(g.s.) at $E_d^{lab}=15$ MeV and
$^{40}$Ca(d,p)$^{41}$Ca(g.s.) at $E_d^{lab}=11$ MeV.
We have repeated this study
using optical potentials fitted specifically to the corresponding elastic
data, using {\sc sfresco} \cite{fresco}. 
The optical potentials resulting from the elastic fits are presented in Table \ref{tab-optpot} and
differ somewhat from the global compilations.
The resulting transfer cross sections, together with the corresponding
elastic scattering, are shown in Figures \ref{c15fitted}, \ref{o17fitted} and  \ref{ca41fitted}.
We will refer to these calculations as DWBAf.
For both the $^{14}$C(d,p) and  the $^{41}$Ca(d,p) reactions, the SF/ANC inconsistency
remains whereas for the  $^{16}$O(d,p) a spectroscopic factor of 0.6-1 produces ANCs
much closer to the value extracted from a sub-Coulomb measurement.
This can be seen from Table \ref{tab-fitted} where SPANCs, SFs and ANCs are
presented as a function of the single particle radius for the three reactions
under study.

\begin{table}
\caption{SFs and ANCs obtain from DWBA analyses:
deuteron optical potential directly fitted to the corresponding elastic data.}
\begin{center}
\begin{tabular}{|c|c|c|c|c|c|c|c|c|c|}
\hline
&\multicolumn{3}{c|}{\nuc{14}{C} (14 MeV)}
&\multicolumn{3}{c|}{\nuc{16}{O} (15 MeV)}
& \multicolumn{3}{c|}{\nuc{40}{Ca} (11 MeV)} \\\hline
$r_0$ (fm) & $b$ (fm$^{-1}$) &  SF  & $C^2$  (fm$^{-1}$)& $b$  (fm$^{-1}$)  &  SF  & $C^2$  (fm$^{-1}$)&
$b$  (fm$^{-1}$) &  SF  & $C^2$  (fm$^{-1}$)\\ \hline
1.01 & 1.34& 1.50 &2.70   &  0.67 &  1.38 & 0.632 &1.32 &  1.23 & 2.159\\
1.11 & 1.37& 1.42 &2.70   &  0.75 & 1.09 & 0.622 &1.66 & 0.985 & 2.729\\
1.21 & 1.41& 1.35 &2.71   &  0.84 & 0.869 & 0.616 &2.09 & 0.748 & 3.272\\
1.31 & 1.45& 1.27 &2.70   &  0.93 & 0.693 & 0.613 &2.62 & 0.545 & 3.754\\
1.41 & 1.49& 1.20 &2.69   &  1.04 & 0.555 & 0.613 &3.28 & 0.387 & 4.172\\
1.51 & 1.54& 1.13 &2.68   &  1.17 & 0.447 & 0.617 &4.10 & 0.271 & 4.557\\
1.61 & 1.58& 1.05 &2.66   &  1.31 & 0.363 & 0.625 &5.11 & 0.189 & 4.945\\
1.71 & 1.63&0.988 &2.64   &  1.46 & 0.297 & 0.636 &6.35 & 0.133 & 5.380\\
1.81 & 1.68&0.921 &2.62   &  1.63 & 0.244 & 0.651 &7.87 & 0.095 & 5.904\\
1.91 & 1.74&0.859 &2.60   &  1.81 & 0.202 & 0.671 &9.75 & 0.069 & 6.555\\ \hline
& \multicolumn{2}{c|}{$C^2_{exp}$} & $1.48\pm0.18$ fm$^{-1}$ & \multicolumn{2}{c|}{$C^2_{exp}$}
& $0.67\pm0.05$ fm$^{-1}$ & \multicolumn{2}{c|}{$C^2_{exp}$}  & $8.36\pm0.42$ fm$^{-1}$ \\
\hline
\end{tabular}
\label{tab-fitted}
\end{center}
\end{table}

\subsection{Adiabatic deuteron potential}
\label{sec-adba}

It is well known that the deuteron breakup can have an effect on the transfer
cross sections. One way to take this into account would be to couple explicitly
the deuteron continuum within the Continuum Discretized Coupled Channel (CDCC) method \cite{cdcc}.
The complexity of the procedure would not be practical for the general experimental community. 
Alternatively, the adiabatic method developed by Johnson and Soper \cite{soper}
can be used to describe the entrance channel and
obtain the transfer cross section while including the deuteron breakup channel. 
This procedure is by far simpler than the CDCC approach and can easily be used in
systematic studies (e.g. \cite{liu,tsang}). 
The three body wavefunction for the $d-T$ system obtained with the adiabatic
approximation  has the correct properties in the range of $V_{np}$ yet should not
be used in the asymptotic region. For this reason we perform finite range calculations
but do not include the remnant term in Eq.(\ref{dwba1}).  Also important is that the deuteron
wavefunction be an eigenstate of $V_{np}$ used in the transfer operator. Transfer matrix
elements with the full RSC potential are not simple.
Thus we have used the central gaussian, refitted to reproduce the correct binding energy 
and the same $D$ as the RSC potential, for calculating both, 
the wavefunction and the transfer operator. 

\begin{figure}[t!]
\resizebox*{0.45\textwidth}{!}{\includegraphics{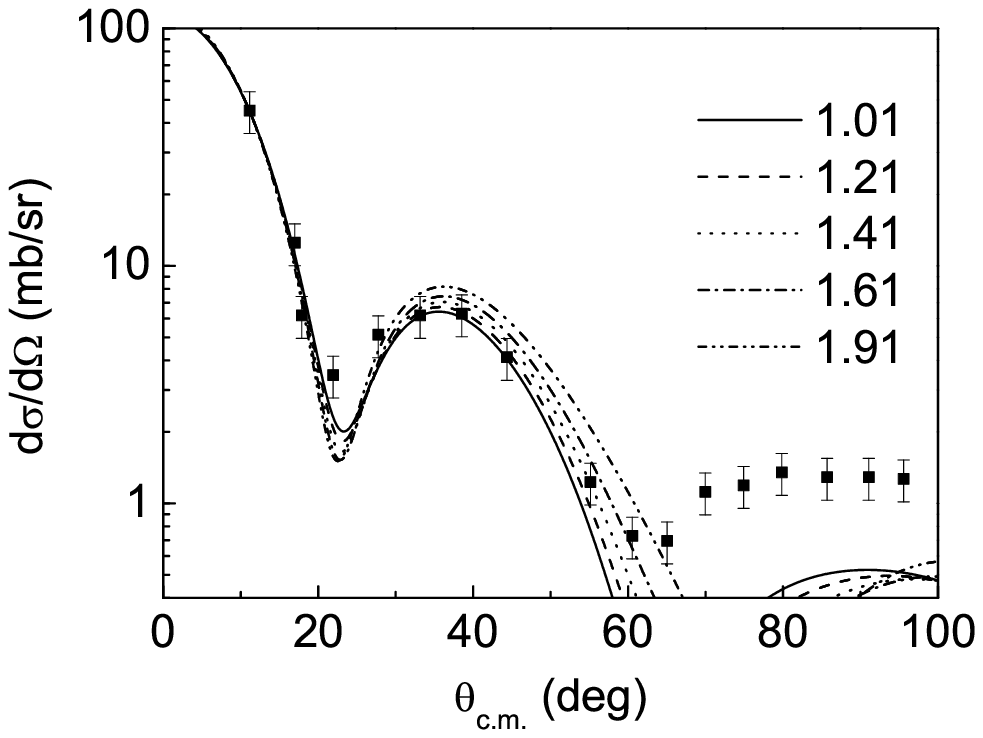}}
\vspace{-0.2cm}
\caption{\label{c15ad}
$\nuc{14}{C}(d,p)\nuc{15}{C}$ at $E_d=14$ MeV with  ADWA.
Data from \cite{goss}.}
\resizebox*{0.45\textwidth}{!}{\includegraphics{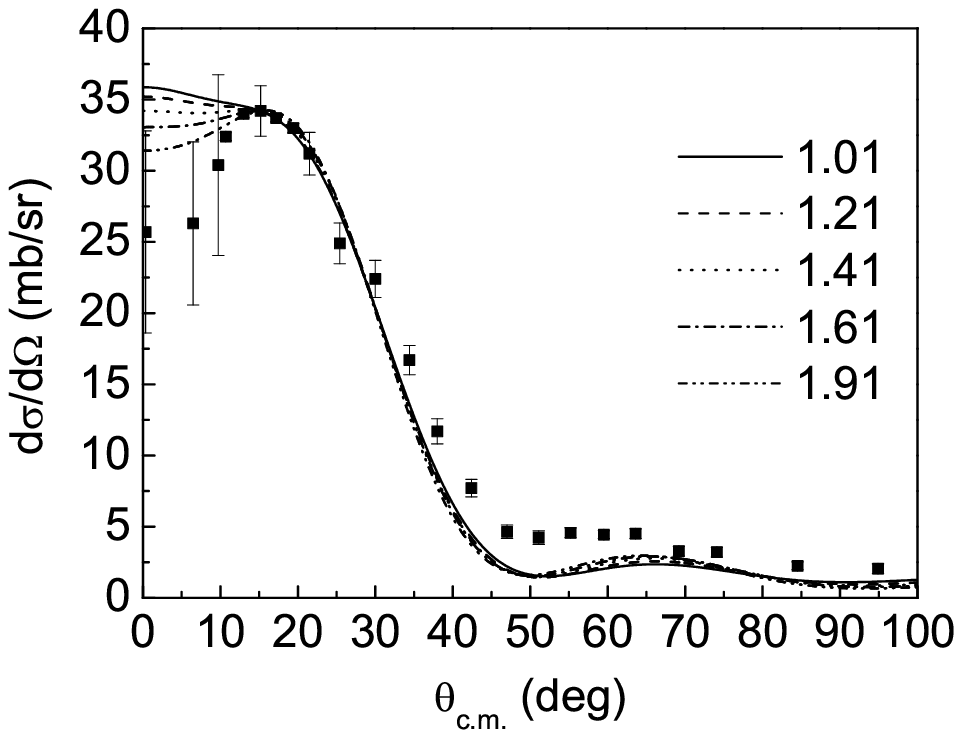}}
\vspace{-0.2cm}
\caption{\label{o17ad}
$\nuc{16}{O}(d,p)\nuc{17}{O}$ at $E_d=15$ MeV with  ADWA.
Data from \cite{eldon}.}
\resizebox*{0.45\textwidth}{!}{\includegraphics{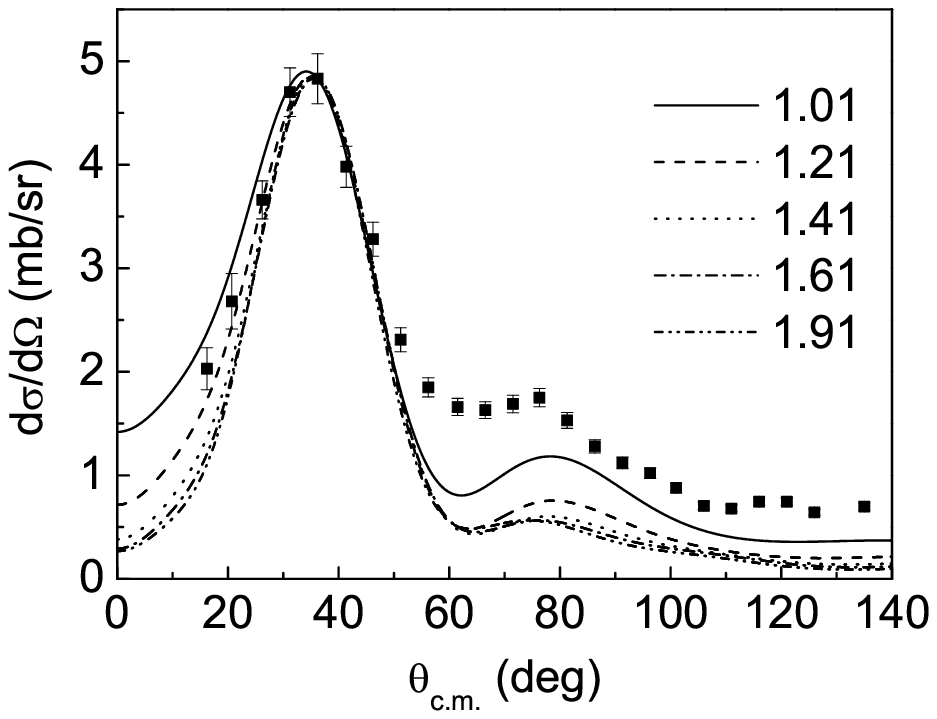}}
\vspace{-0.2cm}
\caption{\label{ca41ad}
$\nuc{40}{Ca}(d,p)\nuc{41}{Ca}$ at $E_d=11$ MeV with  ADWA.
Data from \cite{kocher}.}
\end{figure}

\begin{table}
\caption{SFs and ANCs obtained from ADWA analyses.}
\begin{center}
\begin{tabular}{|c|c|c|c|c|c|c|c|c|c|}
\hline
&\multicolumn{3}{c|}{\nuc{14}{C} (14 MeV)}
&\multicolumn{3}{c|}{\nuc{16}{O} (15 MeV)}
& \multicolumn{3}{c|}{\nuc{40}{Ca} (11 MeV)} \\\hline
$r_0$ (fm) & $b$ (fm$^{-1}$) &  SF  & $C^2$  (fm$^{-1}$)& $b$  (fm$^{-1}$)  &  SF  & $C^2$  (fm$^{-1}$)&
$b$  (fm$^{-1}$) &  SF  & $C^2$  (fm$^{-1}$)\\ \hline
1.01 & 1.34& 1.16 &2.08   &  0.67 &  1.45 & 0.659 &1.32 &  1.80 & 3.14\\
1.11 & 1.37& 1.11 &2.10   &  0.75 & 1.19 & 0.675 &1.66 & 1.30 & 3.57\\
1.21 & 1.41& 1.06 &2.12   &  0.84 & 0.974 & 0.688 &2.09 & 0.895 & 3.92\\
1.31 & 1.45& 1.01 &2.14   &  0.93 & 0.794 & 0.701 &2.62 & 0.609 & 4.19\\
1.41 & 1.49& 0.969 &2.17   &  1.04 & 0.646 & 0.712 &3.28 & 0.410 & 4.43\\
1.51 & 1.54& 0.928 &2.20   &  1.17 & 0.526 & 0.724 &4.10 & 0.278 & 4.67\\
1.61 & 1.58& 0.891 &2.24   &  1.31 & 0.429 & 0.737 &5.11 & 0.189 & 4.94\\
1.71 & 1.63& 0.858 &2.30   &  1.46 & 0.351 & 0.751 &6.35 & 0.131 & 5.28\\
1.81 & 1.68& 0.830 &2.36   &  1.63 & 0.289 & 0.768 &7.87 & 0.092 & 5.72\\
1.91 & 1.74& 0.807 &2.44   &  1.81 & 0.238 & 0.788 &9.75 & 0.066 & 6.30\\ \hline
& \multicolumn{2}{c|}{$C^2_{exp}$} & $1.48\pm0.18$ fm$^{-1}$ & \multicolumn{2}{c|}{$C^2_{exp}$}
& $0.67\pm0.05$ fm$^{-1}$ & \multicolumn{2}{c|}{$C^2_{exp}$}  & $8.36\pm0.42$ fm$^{-1}$ \\
\hline
\end{tabular}
\label{tab-ad}
\end{center}
\end{table}

We use the CH89 nucleon potential for the $U_{nT}$ and $U_{pT}$ at half the deuteron
incident energy and calculate the adiabatic potential using the parametrization
in \cite{wales} which included finite range corrections. 
We will refer to these results as the adiabatic wave approximation
(ADWA) \cite{msu05}. The cross section obtained
for the three reactions under study are presented in Figs. \ref{c15ad}-\ref{ca41ad}.
In table \ref{tab-ad} we show the results for the extracted SF and corresponding ANCs.
An overall reduction of the spectroscopic factors is observed.
The angular distributions show some improvement as compared to those within DWBAg or DWBAf.
However, it becomes clear that deuteron breakup is unable to remove the inconsistency
between SF and ANC, specially in the $^{41}$Ca case, where a persistent factor of 2
remains.

As mentioned above, the ADWA results here presented do not include the remnant.
Results from DWBAg show that remnant contributions
are not important for the lighter cases under study. Even though remnant contributions 
to $^{40}$Ca(d,p)$^{41}$Ca are not negligible, the magnitude is much smaller than
the mismatch observed and we do not expect their inclusion within ADWA would
change the conclusions. Also, the deuteron wavefunction for the adiabatic calculations
is s-wave only (does not contain the d-wave and the tensor interaction). 
This effect alone reduces the spectroscopic factors by $\approx 10$\%. However our conclusions,
namely that deuteron breakup cannot account for the inconsistency
between SF and ANC, remains.

\subsection{Other higher order effects}

The transfer couplings for the examples we are studying are relatively strong, therefore one solution
to the inconsistency between SF and ANC could reside in higher order processes in the
reaction mechanism, other than deuteron breakup. In these calculations our starting point is DWBAf.
First we consider multiple transfer couplings within the
Coupled Reaction Channel approach (CRC).  Note that some of these effects
are accounted for within ADWA. We restrict ourselves to the (d,p) transfer coupling
connecting ground states of the $A$ and $B$. For all three reactions, we increase the number of
iterations  until convergence is achieved. We refit the deuteron
optical potentials, so that the elastic scattering is still well reproduced.

\begin{itemize}
\item For $^{14}$C(d,p)$^{15}$C(g.s.) at $E_d^{lab}=14$ MeV, we find that
there is a significant effect of higher order transfer couplings in the cross section,
which amounts to a reduction of the required SF. Still this does not completely
solve the inconsistency. For example, for $r_0=1.3$ fm, $S=1.013$ and $C^2=2.14$ fm$^{-1}$.

\item
CRC effects for $^{16}$O(d,p)$^{17}$O at 15 MeV are weak (a few percent).

\item
For $^{40}$Ca(d,p)$^{41}$Ca at 11 MeV, CRC increases the transfer cross section
by $\approx 20$\% which produces lower SFs. This makes the inconsistency
SF/ANC more severe.
\end{itemize}

Another source of higher order effects comes
from explicit multi-step excitations of the target.
These targets are spherical but vibrate, and thus there are several
possible couplings that could be considered within
the coupled channel Born approximation (CCBA).
Taking into account the nuclear deformation lengths determined from
nucleon inelastic studies \cite{raman2,kibedi}, we have included the
excited state of the target that couples most strongly to the ground state,  the $3^-$ state.
Again the deuteron optical potential
is refitted to reproduce the elastic scattering correctly. We find that for $^{14}$C, these
couplings have very little effect. This is not the case for both $^{16}$O and $^{40}$Ca.

Let us first consider $^{17}$O.
Its ground state is described in terms of a coupled-channel equation
that generates a dominant $d_{5/2}$ component. Changing $r_0$ can  have a dramatic
effect in the structure composition (mixing in p- and f-waves),
and thus one needs to limit the range of single particle parameters where
a realistic structure is preserved. Results are shown in table \ref{O16_CCBA}. 
Including the coupling to the $3^-$ state,
reduces the transfer cross section, increasing the SF. The resulting ANC of the $d_{5/2}$
for a unit SF, is $0.56$ fm$^{-1}$.

\begin{table}[htbp]
\begin{center}
\caption{Variation of the SFs and the ANCs with $r_0$ for CCBA calculations of
$\nuc{16}{O}(d,p)\nuc{17}{O}$ at $E_d=15$ MeV.}
\label{O16_CCBA}
\begin{tabular}{|c|c|c|c|}
\hline
r0   & SF   &b(1d5/2)  &  $C^2(d5/2)$   \\\hline
1.16 &1.28 &  0.6795  & 0.59   \\
1.20 &1.17 &  0.7003  & 0.58    \\
1.24 &1.08 &  0.7215  & 0.56   \\
1.28 &0.99 &  0.7430  & 0.55  \\
1.32 &0.92 &  0.7647  & 0.53  \\\hline
\end{tabular}
\end{center}
\end{table}
\begin{table}[htbp]
\begin{center}
\caption{Variation of the SFs and the ANCs with $r_0$ for CCBA calculations
of $\nuc{40}{Ca}(d,p)\nuc{41}{Ca}$ at $E_d=11$ MeV. }
\label{Ca11_CCBA}
\begin{tabular}{|c|c|c|c|}
\hline
r0  &  SF   &b(1f7/2) &  $C^2(f7/2)$   \\\hline
1.22& 1.16 & 1.7869  & 2.07  \\
1.24& 1.09 & 1.8541  & 3.75 \\
1.26& 1.03 & 1.9233  & 3.81  \\
1.28& 0.97 & 1.9944  & 3.85 \\
1.30& 0.91 & 2.0674  & 3.88 \\\hline
\end{tabular}
\end{center}
\end{table}

In the $^{41}$Ca case, the situation is different. Taking again only values
of $r_0$ that generate coupled-channel wavefunctions for $^{41}$Ca that are still
dominantly $f_{7/2}$, one obtains transfer cross sections that are smaller than in
the single particle case. Thus the extracted SFs are higher (see table \ref{Ca11_CCBA}). Nevertheless, 
the ANCs associated with the $f_{7/2}$ component are still much smaller than
the number extracted from the sub-Coulomb measurement.
We have performed CRC iterations on CCBA calculations for 
$\nuc{40}{Ca}(d,p)\nuc{41}{Ca}$ at $E_d=11$ MeV.
In this case the transfer couplings involved are: 
$d+^{40}$Ca$(g.s.) \rightarrow p+^{41}$Ca$(g.s.)$ and
$d+^{40}$Ca$(3^-_1) \rightarrow p+^{41}$Ca$(g.s.)$, with spectroscopic factors as in
table \ref{Ca11_CCBA}.
We find the picture does not change significantly.

\subsection{DBWA with a simultaneous fit}
\begin{figure}[t!]
\resizebox*{0.3\textwidth}{!}{\includegraphics{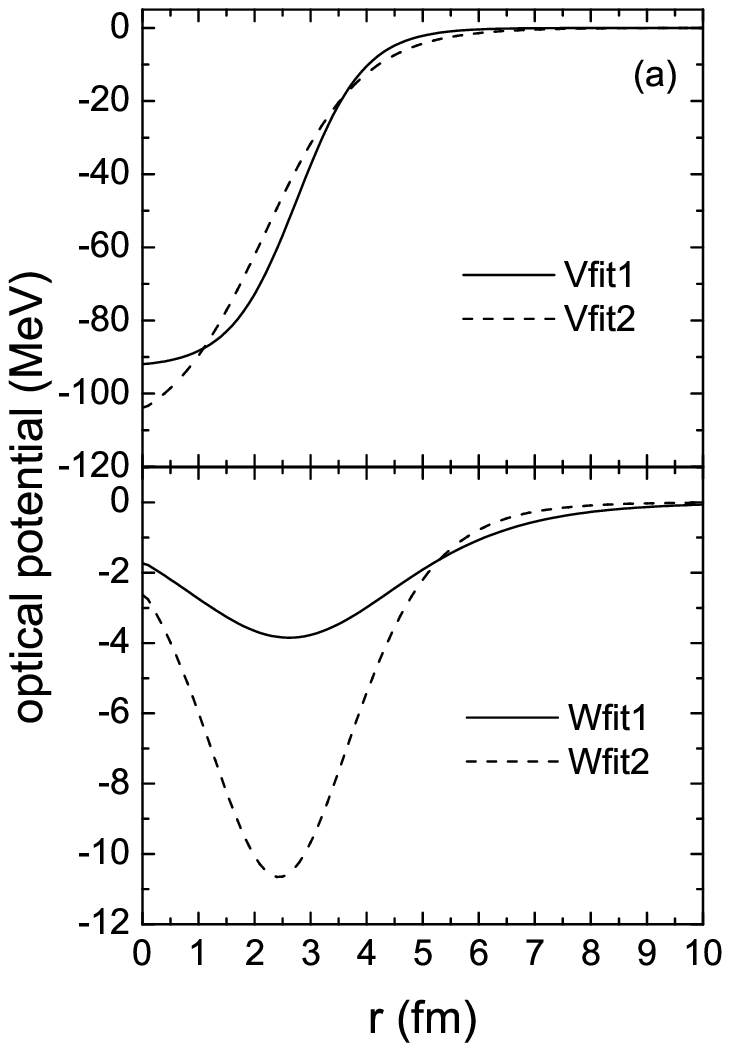}}
\hspace{-0.1cm}
\resizebox*{0.3\textwidth}{!}{\includegraphics{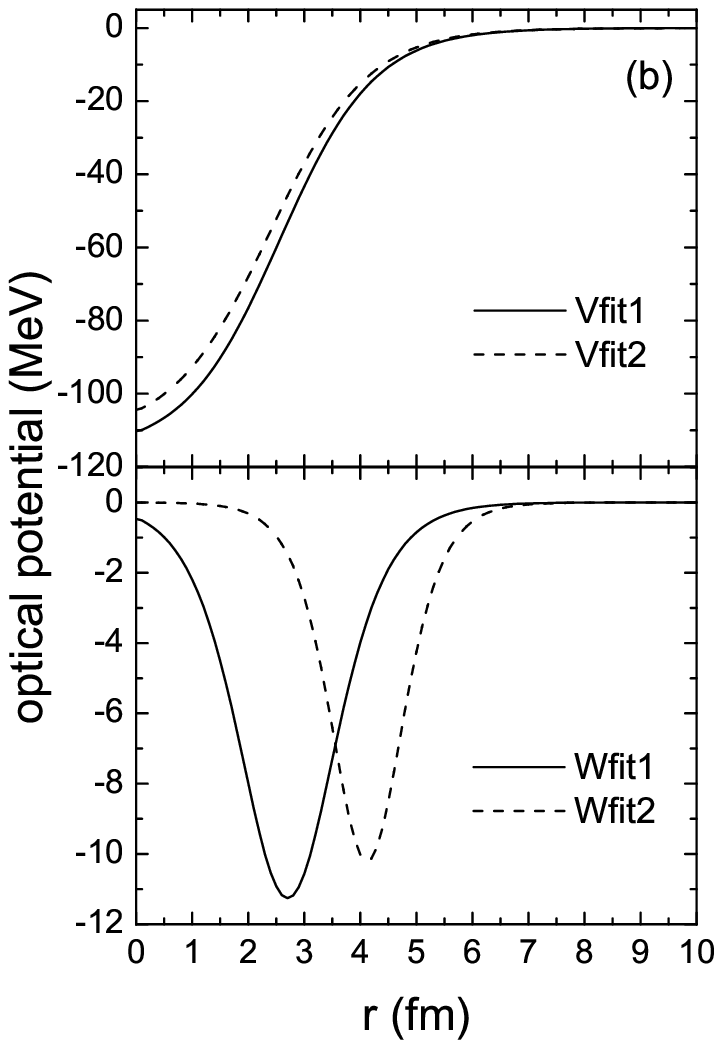}}
\hspace{-0.1cm}
\resizebox*{0.3\textwidth}{!}{\includegraphics{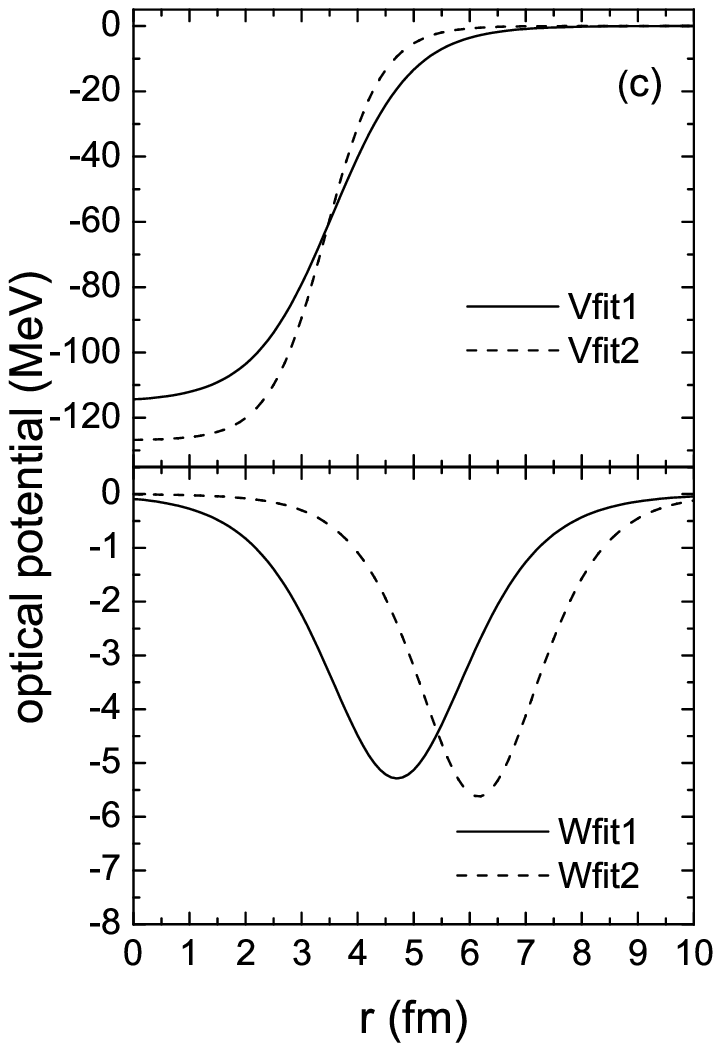}}
\vspace{-0.2cm}
\caption{\label{tran-fit}
Optical potentials: fits to elastic scattering only - real (Vfit1) and imaginary (Wfit1) -
and simultaneous fits to the elastic and transfer channel - real (Vfit2) and imaginary (Wfit2).
a) $\nuc{14}{C}+d$ at $E_d=14$ MeV, b) $\nuc{16}{O}+d$ at $E_d=15$ MeV,
c) $\nuc{41}{Ca}+d$ at $E_d=11$ MeV.}
\end{figure}

\begin{table}[htbp]
\begin{center}
\caption{Single particle orbitals adjusted to produce the
experimental ANC, when SF is unity. }
\label{tab:sp}
\begin{tabular}{|l|c|c|c|}
\hline
        &  $^{15}$C   & $^{17}$O &  $^{41}$Ca   \\\hline
$r_0$ (fm) & 1.10 & 1.25  & 1.30  \\
$a$ (fm)   & 0.52 & 0.60  & 0.75 \\
$C^2$ (fm$^{-1})$& 1.62 & 0.68  & 8.24  \\
$R_{rms}(fm) $& 5.1 & 3.5  & 4.2 \\
\hline
\end{tabular}
\end{center}
\end{table}
It is impossible to rule out other excitation mechanisms which may contribute to the
transfer process, however we do not expect these will be stronger than the ones discussed
in  section IIID. One interesting question is whether there would be any way
of obtaining the desired consistency between SF and ANC within the simplified
DWBA picture, specially in the $^{41}$Ca case, where the mismatch is so large. 
We assume that we can introduce all higher order processes into
a local effective deuteron optical potential. We impose that the SF for a neutron
outside a closed shell be unity, and chose the single particle parameters such
that they reproduce the ANC extracted from experiment. We perform a nine parameter
fit of the deuteron optical potential to both the elastic and the transfer data.
Our starting potential is that obtained in section \ref{sec-dwbaf}.
The fit consists of a standard $\chi^2$ minimization procedure and the code {\sc sfresco}
is used \cite{fresco}. 
For each case we do find a potential that is able to reproduce the elastic
and the transfer simultaneously, under the constraint of SFs and ANCs.
The resulting real (Vfit2) and imaginary (Wfit2) potentials from this fit are presented for
all three reactions, as
the dashed lines ({\it fit2}) in Figs. \ref{tran-fit}. The parameters are explicitly shown
in table \ref{tab-optpot2}. The solid lines correspond to the previous elastic fit ({\it fit1}),
whose parameters were shown in table \ref{tab-optpot}.
It appears that for $\nuc{14}{C}+d$ at $E_d=14$ MeV, the real part becomes slightly more
diffuse, whereas for $\nuc{41}{Ca}+d$ at $E_d=11$ MeV, it becomes clearly
less diffuse. The $\nuc{16}{O}+d$ at $E_d=15$ MeV case is rather unchanged, which
reflects the fact that to start with the inconsistency in this case was minor.
We find that the most significant change in the resulting potentials is 
an increase of the imaginary part in the surface region, which is obtained either by a large
imaginary depth or by a shift toward the surface. This suggests that the standard DWBA is 
overestimating the surface contribution.

\begin{table}
\caption{Deuteron optical potential parameters resulting from the simultaneous fit to elastic
and transfer data.}
\begin{center}
\begin{tabular}{|c|c|c|c|c|c|c|c|c|c|c|}
\hline
$A$& $E_{beam}$ (MeV) & $V_{R} $(MeV)  &  $R_{R}$ (fm)  & $a_{R}$ (fm) & $W_{d}$(MeV)
& $R_I$ (fm)  &  $a_I$ (fm)  & $V_{so}$(MeV) & $R_{so}$ (fm) & $a_{so}$ (fm)\\\hline
14 & 14 & 113.015 & 0.902 & 0.875 & 10.659 & 1.007 & 0.900 & 14.342  & 1.281 &  0.694 \\
16 & 15 & 111.531 & 0.948 & 0.871 & 10.211 & 1.635 & 0.437 & 10.991  & 0.505 &  0.541 \\
40 & 11 & 126.922 & 1.004 & 0.500 &  5.625 & 1.800 & 0.736 &  6.606  & 0.988 &  0.4038 \\\hline
\end{tabular}
\label{tab-optpot2}
\end{center}
\end{table}

\begin{figure}[t!]
\resizebox*{0.7\textwidth}{!}{\includegraphics{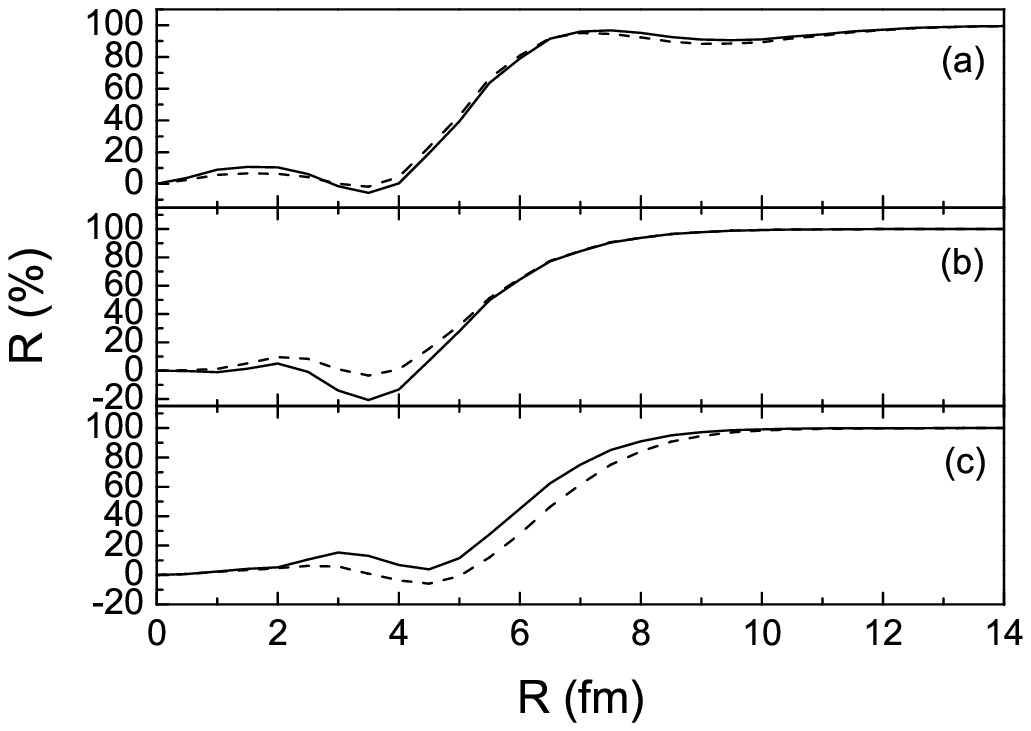}}\\
\vspace{-0.2cm}
\caption{\label{ratio2}
Peripherality test for the reactions (using a cut in the radial
distance between projectile and target):
a) $\nuc{14}{C}+d$ at $E_d=14$ MeV, b) $\nuc{16}{O}+d$ at $E_d=15$ MeV,
c) $\nuc{41}{Ca}+d$ at $E_d=11$ MeV.}
\end{figure}

Changes in the deep interior do not reflect sensitivity of the transfer process to this
region. Transfer reactions are known to be surface peaked. This usually means that impact
parameters smaller than the sum of the projectile and target radii do not contribute
to the reaction. In $A(d,p)B$, this is related to the cutoff on $\bf R_{dA}$. The results of
this standard peripherality check is shown in Fig. \ref{ratio2} for the three reactions
under study. The percentage ratio of the cross section integrated up to $R_{dA}=R$
to the total transfer cross section R(\%) is plotted. In the deep interior this ratio
is zero and it goes to $100$ \% for large distances. The region where it increases 
rapidly corresponds to the surface region, where the transfer takes place.

\begin{figure}[t!]
\resizebox*{0.7\textwidth}{!}{\includegraphics{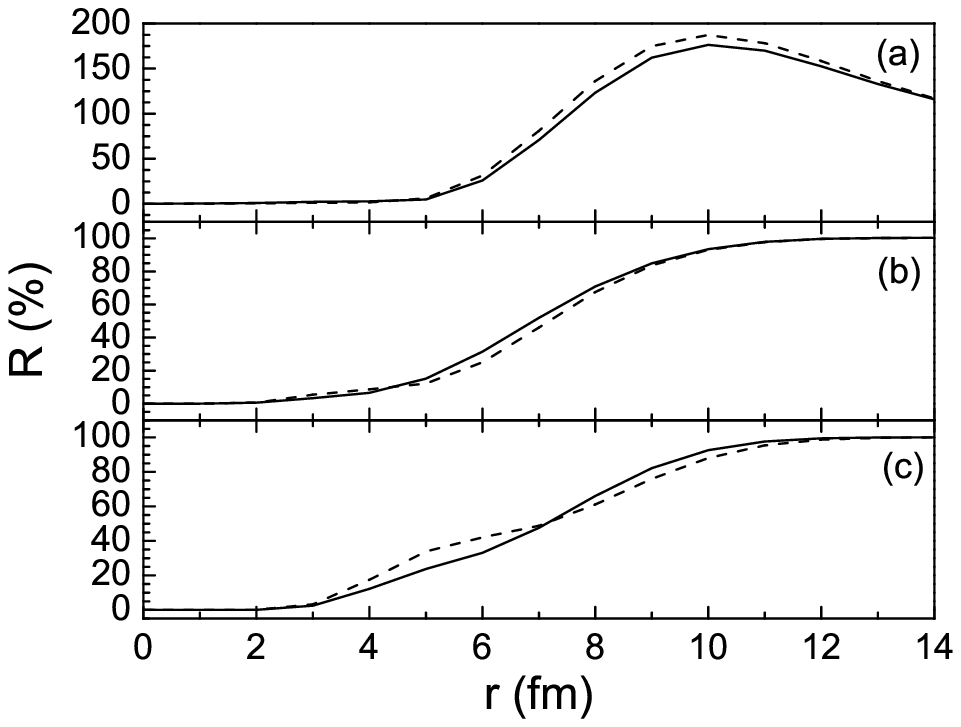}}\\
\vspace{-0.2cm}
\caption{\label{ratio}
Peripherality test for the reactions (using cutoff in the radial
distance $R_{nA}$):
a) $\nuc{14}{C}+d$ at $E_d=14$ MeV, b) $\nuc{16}{O}+d$ at $E_d=15$ MeV,
c) $\nuc{41}{Ca}+d$ at $E_d=11$ MeV.}
\end{figure}

However, to probe the sensitivity to the SF, one needs to analyse the dependence
with $\bf R_{nA}$. Under a zero range approximation of the deuteron, these two tests would
be identical. When taking into account the finite range of $V_{np}$ and the remnant
part of the full transition operator, 
peripherality in $\bf R_{dA}$ is not equivalent to peripherality in $\bf R_{nA}$.
To illustrate this fact, peripherality tests
were performed by evaluating the interior contribution to the total transfer cross section
relative to $\bf R_{nA}$. These peripherality tests are based on DWBAf {(\it fit 1)}
and DWBA with the deuteron potential fit simulaneously to elastic and transfer (fit 2).
We take the radial integrals in
the coordinates of $R_{nA},R_{pB}$, and truncate the integration in $R_{nA}$
to a maximum value r:
\begin{equation}
\frac{d \sigma}{d \Omega}(r) \sim
(\int_0^\infty \int_0^r \; \psi_{f}^{(-)} I^{B}_{An} \Delta V \varphi_{pn}\,\psi_{i}^{(+)}
{\bf d R_{nA}} {\bf d R_{Bp}})^2.
\label{cuts}
\end{equation}
The percentage ratio
of this value to the full integration is presented in Fig. \ref{ratio}.
As before, if $r=0$ fm, the ratio should be zero, and it should tail off at 100 \%
when r becomes very large. Results show that, in all cases there is no contribution up to 2 fm.
One can also see that, even
though 14 MeV is above the Coulomb barrier, the $\nuc{14}{C}(d,p)$ happens at rather
large distances, due to its loosely bound nature.
Contributions up to $20$ \% from the surface/interior are present in both
$\nuc{16}{O}+d$ at $E_d=15$ MeV and  $\nuc{41}{Ca}+d$ at $E_d=11$ MeV.

\section{Discussion and Conclusions}

In this work we have studied transfer reactions to states considered good single particles,
with three different Q-values,
namely  $^{14}$C(d,p)$^{15}$C(g.s.) at $E_d^{lab}=14$ MeV,
$^{16}$O(d,p)$^{17}$O(g.s.) at $E_d^{lab}=15$ MeV
$^{40}$Ca(d,p)$^{41}$Ca(g.s.) at $E_d^{lab}=11$ MeV.
All these reactions are above the Coulomb barrier and therefore
contain some information from the interior.
The standard DWBA method, using global optical potentials and the typical single
particle parameters, produces SFs in agreement with shell model predictions,
however the corresponding ANCs are not consistent with those extracted from
independent measurements. If one imposes, within the DWBA formulation,
ANCs that are consistent with the experimental values, the extracted SFs are
significantly reduced compared to the shell model predictions.

Some improvements on the SF/ANC mismatch can be obtained by using a deuteron
optical potential fitted directly to the corresponding elastic data, at the
relevant energy. In particular, for $^{16}$O(d,p) we obtain SF/ANC consistency.
However the problem for the other two cases is not resolved.
The deuteron adiabatic potential, which takes into account breakup, can change
the SF up to $30$ \%. This improves the situation for $^{14}$C(d,p) but fails to bring
the $^{41}$Ca ANC anywhere close to the value extracted from an independent
sub-Coulomb measurement. 

Deuteron breakup is not the only important ingredient to the solution of this problem: 
in all cases we find an influence of surface higher order
effects. For $^{14}$C(d,p)$^{15}$C(g.s.) at $E_d^{lab}=14$ MeV, CCBA effects
are negligible but CRC effects produce a reduction of the ANC consistent
with unity SF: $C^2=2.14$ fm$^{-1}$, a significant reduction compared to the
DWBA case, which together with the reduction from deuteron breakup produces
an ANC much closer to the experimental value $C^2=1.48 \pm 0.18$ fm$^{-1}$.
For $^{16}$O(d,p)$^{17}$O(g.s.) at $E_d^{lab}=15$ MeV, CRC effects are small but
CCBA couplings to the $3^-$ state in $^{16}$O were found to be important.
This brings down the $C^2$ from 0.76 to 0.56, neither too far from
the experimental number $0.67 \pm 0.05$ fm$^{-1}$.
Finally, the $^{40}$Ca(d,p)$^{41}$Ca(g.s.) case remains problematic: both
CRC and CCBA are relevant but act in the wrong direction.
Overall, an ANC, consistent with unity SF, falls short
by nearly a factor of 2.


Contrary to (e,e'p) measurements, transfer reactions are surface peaked and
it is disconcerting that the traditional methods to handle higher order effects
at the surface are not able to solve the SF/ANC discrepancy for one of our test cases.
The very fact that, even when the energies are well above the Coulomb barrier,
there is such a large contribution from the peripheral region, makes it extremely
important to pin down the ANC input unambiguously. We cannot rule out the possibility
of a problem in the  $^{41}$Ca ANC we extracted from other data.
Experiments to measure ANCs for this case is crucial to settle this matter.
In the future we suggest that experiments be designed for the extraction
of ANCs in parallel with the corresponding experiments aimed at  extracting  SFs.

\section*{Acknowledgements}

This work was supported by the National Science Foundation, under grant PHY-0456656
and the U.S. Department of Energy, under Grant No. DE-FG02-93ER40773. D.Y. Pang acknowledges
the support of the National Superconducting Cyclotron Laboratory during his visit.


\begin{thebibliography}{sfanc}
\bibitem{brown} B.A. Brown {\it et al.}, Phys. Rev. C 65 (2002) 061601(R).
\bibitem{gfmc} R. B. Wiringa, Presentation at the Workshop on {\em
New perspectives on p-shell nuclei - the nuclear shell model and beyond},
Michigan State University on July 22-24, 2004 (http://www.nscl.msu.edu/~brown/p-shell-2004/pdf/wiringa.pdf).
\bibitem{gfmc-eep} L. Lapik\'as, J. Wesseling, and R. B. Wiringa, Phys. Rev. Lett. 82, 004404 (1999).
\bibitem{goncharov}  S. A. Goncharov {\it et al.},  Sov. J. Nucl. Phys. 35, 383 (1982).
\bibitem{austern70} N. Austern, Direct Nuclear Reaction Theories (Wiley, New York, 1970).
\bibitem{jpg-rev}   Jim Al-Khalili and Filomena Nunes,
        J. Phys. G: Nucl. Part. Phys. 29 (2003) R89.
\bibitem{kramer} G.J. Kramer, H.P. Blok and L. Lapikas,
    Nucl. Phys. A 679 (2001) 267.
\bibitem{barbieri} C. Barbieri and L. Lapik\'as,  Phys. Rev. C 70, 054612 (2004).
\bibitem{knock} A. Navin {\it et al.}, Phys. Rev. Lett. 81 (1998) 5089.
\bibitem{gade} A. Gade et al., Phys. Rev. Lett. 93, 042501 (2004).
\bibitem{gregers} P. G. Hansen and J. A. Tostevin, Annu. Rev. Nucl. Part. Sci. 53, 219 (2003).
\bibitem{born} R.J. Furnstahl, H.-W. Hammer, Phys.Lett. B531 (2002) 203-208.
\bibitem{schiffer} J. P. Schiffer {\it et al.},
    Phys. Rev. 164, 1274 (1967).
\bibitem{dwba} Xiaodong Tang et al., Phys. Rev. C 69 (2004) 055807.
\bibitem{iliadis04}  C. Iliadis, M. Wiescher, Phys. Rev. {\bf C 69}, 064305 (2004).
\bibitem{liu} X.D. Liu {\it et al.},
    Phys. Rev. C 69 (2004) 064313.
\bibitem{tsang} M.B. Tsang, Jenny Lee, W.G. Lynch, Phys. Rev. Lett. 95, 222501 (2005).
\bibitem{delaunay} F. Delaunay, F. M. Nunes, W. G. Lynch, and M. B. Tsang,
    Phys. Rev. C 72, 014610 (2005).
\bibitem{jenny} Jenny Lee {\it et al.}, Phys. Rev. C 73, 044608 (2006).
\bibitem{us} A.M. Mukhamedzhanov and F.M. Nunes, Phys. Rev. C 72, 017602 (2005).
\bibitem{o17anc1}  M.A. Franey et al., Nucl. Phys. A324, 193 (1979).
\bibitem{o17anc2}  S. Burzynski et al., Nucl. Phys. A399, 230 (1983).
\bibitem{sauvan} E. Sauvan {\it et al.},  Phys. Rev. C 69, 044603 (2004).
\bibitem{maddalena} V. Maddalena {\it et al.}, Nucl. Phys. A 682 (2001) 332c-338c.
\bibitem{c15anc} Progress report 2001-2002 p.1-6, Texas A\&M.
\bibitem{timofeyuk} N.K. Timofeyuk, D. Baye, P. Descouvemont, R. Kamouni and I.J. Thompson,
    Phys. Rev. Lett. 96, 162501 (2006).
\bibitem{ca41anc} H. Sch\"ar, D. Trautmann, and E. Baumgartner, Helv. Phys. Acta 50, 29 (1977).	
\bibitem{ch89} R. L. Varner {\it et al.}, Phys. Rep. 201, 57 (1991).
\bibitem{soper} R.C. Johnson and P.J.R. Soper, Phys. Rev. C 1 (1970) 055807.
\bibitem{rsc} V. Reid, Ann. Phys. (N.Y.) 50, 411 (1968).
\bibitem{perey} C. M. Perey and F. G. Perey, At. Data Nucl. Data Tables 17, 1 (1976).
\bibitem{goss} J.D. Goss et al., Phys. Rev. C 12, 1730 (1975).
\bibitem{eldon} Eldon L. Keller, Phys. Rev. 121, 820 (1961).
\bibitem{kocher} D.C. Kocher and W. haeberli, Nucl. Phys. A172, 652(1971).
\bibitem{database} http://groups.nscl.msu.edu/nscl\_library/pddp/database.html
\bibitem{fresco} I.~J. Thompson, Comput.\ Phys.\ Rep.\ {\bf 7}, 3 (1988).
\bibitem{o16el} C. E. Busch {\it et al.}, Nucl. Phys. A223 (1974) 183.
\bibitem{ca40el} C.C. Foster, Phys. Rev. 181 (1960) 1529.
\bibitem{cdcc} Y. Sakuragi, M. Yahiro, and M. Kamimura,
    Prog.\ Theor.\ Phys.\ Suppl. {\bf 89},  136  (1986).
\bibitem{wales} G.L. Wales and R.C. Johnson, Nucl. Phys. A 274 (1976) 168.
\bibitem{msu05}
R.C. Johnson,  in {\em Proceedings of the Second Argonne/MSU/JINA/INT RIA
  Workshop on Reaction Mechanisms for Rare Isotope Beams}, edited by B.~A.
  Brown (AIP 791 (2005) 132).
\bibitem{raman2}S. Raman, C. W. Nestor, and P. Tikkanen; Atomic Data and Nucl. Data Tables, 78, 1 (2001).
\bibitem{kibedi}T. Kibedi and R. H. Spear; Atomic Data and Nucl. Data Tables, 80, 35 (2002).




\end{thebibliography}
\end{document}